\documentclass[10pt,pre,twocolumn,a4paper,showpacs]{revtex4}
\usepackage{graphicx}
\usepackage{amssymb}
\usepackage{amsmath}

\begin{document}

\title{Aggregation in a mixture of Brownian and ballistic wandering particles}

\author{S. G. Alves} 
\affiliation{Departamento de F\'{\i}sica, Universidade Federal de Minas
Gerais, CP 702, 30161-970, Belo Horizonte, MG, Brazil}

\author{S. C. Ferreira Jr.}
\affiliation{Departamento de F\'{\i}sica, Universidade Federal Vi\c{c}osa, 36571-000, Vi\c{c}osa, MG, Brazil}

\date{\today}

\begin{abstract}
In this paper, we analyze the scaling properties of a model that has as limiting cases the diffusion-limited aggregation (DLA) and the ballistic aggregation (BA) models. This model allows us to control the radial and angular scaling of the patterns, as well as, their gap distributions. The particles added to the cluster can follow either ballistic trajectories, with probability $P_{ba}$, or random ones, with probability $P_{rw}=1-P_{ba}$. The patterns were characterized through several quantities, including those related to the radial and angular scaling. The fractal dimension as a function of $P_{ba}$ continuously increases from $d_f\approx 1.72$ (DLA dimensionality) for $P_{ba}=0$ to $d_f\approx 2$ (BA dimensionality) for $P_{ba}=1$. However, the lacunarity and the active zone width exhibt a distinct behavior: they are convex functions of $P_{ba}$ with a maximum at $P_{ba}\approx1/2$. Through the analysis of the angular correlation function, we found that the difference between the radial and angular exponents decreases continuously with increasing $P_{ba}$ and rapidly vanishes for $P_{ba}>1/2$, in agreement with recent results concerning the asymptotic scaling of DLA clusters.
\end{abstract}

%\keywords{}

\maketitle

\section{Introduction}
Nonequilibrium growth models of clusters of identical particles have attracted a lot of attention in the last two decades \cite{Meakinbook,Sander}. Certainly, the most studied one is the diffusion-limited aggregation (DLA) model proposed by Witten and Sander \cite{Witten} for the modeling of metal particle aggregation. In spite of its simplicity, this model is a noteworthy example in which a very simple algorithm generates disordered patterns with nontrivial scaling \cite{Somfai, Mandelbrot, Jensen, Goold}. In the DLA model, the particles are released at points distant from the cluster and follow random walks of unitary steps, which are stopped when these particles contact the cluster and irreversibly aggregate to it. Another important aggregation model with nontrivial scaling is the ballistic aggregation (BA) \cite{Vold,Liang}, in which ballistic trajectories at random directions replace the random walks characteristic of the DLA model. Differently from the DLA case, the patterns generated by the BA model are asymptotically homogeneous (fractal dimension equal to the space dimension), but this asymptotic regime is characterized by a power law approach \cite{Liang,Ferreira2005}.

Several generalized versions of the DLA model were proposed \cite{Meakinbook}, including those concerned with the transition between DLA and BA models \cite{Ferreira2005,  Castro, Kim}. In a recent work, Ferreira \textit{et al.} \cite{Ferreira2005} built a model in which the particles follow random trajectories with drifts at random directions. The bias is controlled by a parameter $\lambda$ that, when varied, leads the model from BA ($\lambda=0$) to DLA ($\lambda$=1). The clusters are asymptotically non-fractal (exhibiting the BA-like scaling with fractal dimension $d_f=2$) for any $\lambda\ne 1$, but following a DLA-like scaling ($d_f\approx 1.71$) for short length scales. Using the concept of scaling functions, these ideas can be summarized as
\begin{equation}
  M(l)= l^{d_{DLA}}f(l/\xi),
\label{eqM}
\end{equation}
in which
\begin{equation}
f(x)\sim\left\{ \begin{array}{l}
        const. \mbox{~if~} x\ll 1\\
        x^{d-d_{DLA}} \mbox{~if~} x\gg 1.
        \end{array}\right.
\label{eqf}
\end{equation}
In these equations, $l$ is the size of the cluster, $d$ is the space dimension, $d_{DLA}$ is the DLA fractal dimension, and $\xi$ is the characteristic crossover radius from DLA- to BA-like scaling regimes. The last diverges as $\xi\sim|1-\lambda|^{-\nu}$ where $\nu=0.61\pm0.01$. This transition between DLA and BA regimes was introduced by Meakin \cite{Meakin1983} in a early generalization of the DLA model. This model is very similar to that analyzed by Ferreira \textit{et al.} \cite{Ferreira2005}. The main difference is that the drift of all trajectories is fixed at a lattice direction in the Meakin's model, in which, along the trajectory, the particle is moved one step in the drift direction with probability $P$, or moves at random to one of its four nearest-neighbor sites with probability $1-P$. The model generates patterns with a growth tendency along the opposite direction of the drift. In this work, Meakin argued that the crossover radius scales as $\xi=P^{-1}$ while Nagatani \cite{Nagatani}, through a real-space renormalization group approach, found $\xi\sim P^{-1/(d-d_f)}$.

Early studies indicated that the DLA aggregates exhibit distinct scaling exponents for radial and angular correlations. The two-point correlation function $\Gamma$, which gives the probability to find a particle at a distance between $r$ and $r+\delta r$, $\delta r\ll r$,  from another particle, is given by \cite{Meakinbook}
\begin{equation}
\Gamma(r)\sim r^{-\alpha_r},
\end{equation}
where $\alpha_r=d-d_f$. For off-lattice DLA, the exponent found was $\alpha_r\approx 0.29$. In turn, the angular correlation function $\Phi_R$, which gives the probability to find two particles with angular separation $\theta$ measured in relation to the origin at a distance $R$, follows a power law decay for small $\theta$
\begin{equation}
\Phi_R(\theta)\sim \theta^{-\alpha_\theta}.
\end{equation}
For uniform self-similar fractals $\alpha_r=\alpha_\theta$, but this was not observed for the DLA model, for which $\alpha_{\theta}\approx 0.41$ \cite{Meakin1985}.

In this work, we used careful simulations to analyze the transition between DLA- and BA-like regimes in a model where both random and ballistic trajectories are included. Differently from the previously described models \cite{Meakin1983,Ferreira2005}, the trajectories are not characterized by distinct behaviors at short (random) and large (ballistic) scales. Indeed, we are interested in the scaling of clusters grown from a mixture of particles, which can follow either ballistic or random trajectories. The paper outline is the following. In section \ref{sec:model}, the algorithm and the computer implementations are described. In section \ref{sec:scaling}, the quantities describing the scaling of the cluster are introduced and the corresponding results discussed. Finally, we summarize the results of the paper in section \ref{sec:conclusions}.

\section{Model and computer implementation}

\label{sec:model}

The present model is concerned with the aggregation of particles which can follow either ballistic or random trajectories. Notice that there are two kinds of independent trajectories, which do not exhibit both random and ballistic components simultaneously, as those used in reference \cite{Ferreira2005}. Two-dimensional off-lattice simulations starting from a single particle of unitary radius stuck to {the origin were considered. The model's rules are very simple. For each time step, a kind of trajectory, Brownian or ballistic, is chosen with probabilities $P_{rw}$ and $P_{ba}$, respectively. Then, particles are launched one at a time using the selected rule until one of them sticks to the cluster}. Notice that in these rules we are {taking in to account} only particles, which are added to the cluster and not those that get too far from the cluster and are excluded.

\begin{figure}[hbt]
\begin{center}
\includegraphics[width=8.0cm,height=!]{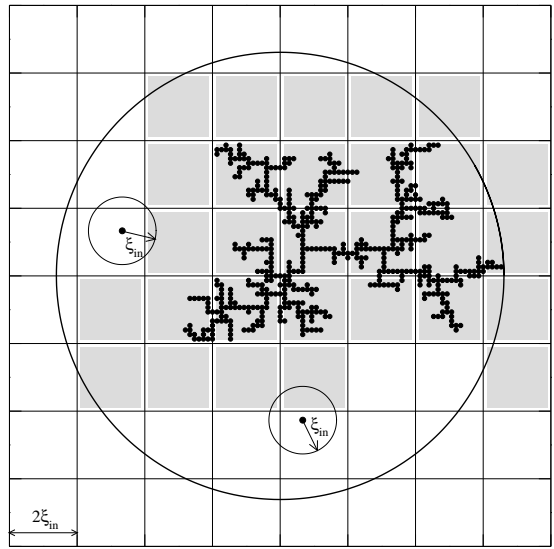}
\vspace{-0.5cm}
\caption{Schematic representation of the ``optimized random trajectories''. We show a DLA aggregate and a mesh of cells $2\xi_{in}\times2\xi_{in}$. Long steps are forbidden in the gray boxes and allowed in the white ones. Also, two long steps are illustrated.}
\label{fig:otm_in}
\end{center}
\end{figure}

The particles are released at a launching circle of radius $r_l>r_{max}$ centered at the origin, where $r_{max}$ is the largest distance from the origin among all the aggregated particles. As in the off-lattice DLA model \cite{Tolman}, the cluster grows when a walker contacts any particle of the aggregate and sticks to the corresponding position. Also, if the walker crosses a killing circle of radius $r_k > r_l$ centered at the origin, it is discarded and a second one released at a new random point on the launching circle. The previously introduced variables ($r_l$ and $r_k$) should be as large as possible. However, computational limitations restrict their values. For sake of convenience, we introduce a new variable $\delta$ by the definition $r_l=r_{max}+\delta$. The choices of $\delta$ and $r_k$ depend on the kind of trajectory. For random walks, $r_l$ can be a few particle diameters larger than $r_{max}$ whereas $r_k$ must be very large \cite{Meakinbook}. Consequently, we used $\delta=5$ and $r_k=100r_{max}$. For ballistic trajectories undesirable shadow effects are reduced as $r_l$ increases \cite{Tang,Yu}. Thus, we used $\delta=100r_{max}+800$. Finally, ballistic trajectories which crosses the launching circle can not turn back inside the circle and, so, we chose $r_k=r_l+2$.

Large scale off-lattice simulations with rigorous statistical sampling require very efficient algorithms. Firstly, it is necessary to improve the efficiency of the routines for the trajectories. In the case of random walks, we used a standard method, in which particles outside the launching circle take long steps $\xi_{out}$ if these steps do not bring up a particle inside the launching circle. An adequate choice is $\xi_{out}=\max(r-r_{max}-5,1)$, where $r$ is the distance of the walker from the origin. Also, in the inner region of the circle delimiting the cluster there are large empty regions. So, we adopted an algorithm like those used by Ball and Brady \cite{Ball}, in which particles inside the launching circle can take a long step of length $\xi_{in}$ if they  cannot cross any part of the aggregate along this long step. 

An efficient determination of $\xi_{in}$ is decisive in order to guarantee the success of the procedure. In order to accomplish this, we define a coarse-grained mesh with cells of size $2\xi_{in}\times 2\xi_{in}$ as illustrated in figure \ref{fig:otm_in}. The boxes depicted in gray are those in which the random walk could reach the cluster after a step of length $\xi_{in}$ and, consequently, a long step is forbidden. In this case, we have two options: the particle executes a unitary step or tries a shorter step of length $\xi_{in}'$, where $1<\xi_{in}'<\xi_{in}$, using another auxiliary coarse-grained mesh. Indeed, we can use several auxiliary meshes in order to maximize the algorithm efficiency. In this work, three values, $\xi_{in}=8,16$ and $32$, were used. Through this procedure, large simulations become up to two orders of magnitude faster than those using only long steps outside the launching circle.

Finally, an efficient search mechanism for determining when the walker has or not a contact with the aggregate is indispensable. We used two procedures for this task. In the first one, the particle positions are mapped on a square lattice by the approximation of their real coordinates to the nearest integer, producing an on-lattice auxiliary cluster. So, the verification of the contact is done only if the particle is inside a nearest or a next-nearest empty site neighboring the on-lattice auxiliary cluster. In the second optimization procedure, a coarse-grained mesh of cells $8\times 8$ was used to limit the verification on a region near the walker position.

\section{\label{sec:scaling} Radial and angular scaling}

\begin{figure}[h]
\begin{center}
\includegraphics[width=8.5cm,height=!]{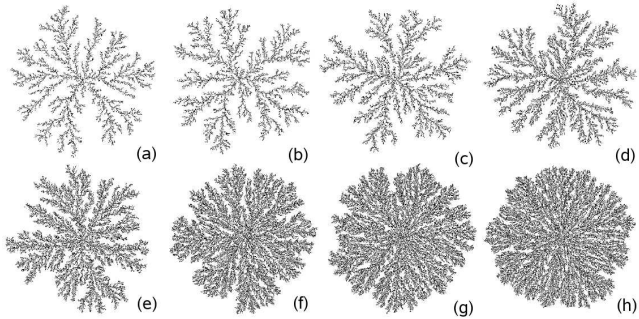}
\end{center}
\vspace{-0.5cm}
\caption{\label{fig:pads} Growth patterns for distinct values of the parameter $P_{ba}$: (a) $0.1$, (b) $0.3$, (c) $0.5$, (d) $0.7$, (e) $0.9$, (f) $0.95$, and (h) $0.99$. All simulations were stopped when the cluster reaches the border of a region $750\times750$ centered at the origin. The number of particles varies from $N\simeq 3\times10^4$ ($P_{ba}=0.1$) to $N \simeq 1.6\times10^5$ ($P_{ba}=0.99$).}
\end{figure}

The model exhibits a smooth transition from DLA to BA-like growth patterns as the parameter $P_{ba}$ varies from $0$ to $1$. In figure \ref{fig:pads}, in which growth patterns for distinct probabilities $P_{ba}$ are shown, one can observe clusters that are qualitatively very similar to the DLA aggregates for $P_{ba}\le 0.5$, whereas BA-like clusters are observed for $P_{ba}\ge0.95$. These patterns were grown until a part of them reaches the edge of a region of size $L\times L$, where $L=750$. Notice that $P_{ba}=0.5$ is an important case, in which a half of the particles were grown using DLA model and the another half were grown using the BA model.

The scaling properties of the clusters were characterized by the radial and angular distributions of the particles, as well as, the size of their active (growing) zone. 
We grew 2d off-clusters limited by a region of size $10^4\times10^4$. Again, the simulations were stopped when the cluster reaches border of this region. The radial scaling was characterized by the fractal dimension and the lacunarity \cite{Meakinbook}. The fractal dimension $d_f$ and the lacunarity $A$ were determined using the mass-radius and the radius of gyration methods. The mass-radius method consists of counting the number of particles $M$ in a hypersphere of radius $r$ centered at the origin. We expect a power law relation for self-similar clusters
\begin{equation}
M(r)=A_mr^{d_f},
\end{equation}
where the amplitude $A_m$ is a measure of the lacunarity. To be more precise, the DLA clusters are characterized by a multiscaling, which leads to a slow convergence to the asymptotic scaling exponents. However, large scale simulations confirm that these clusters are asymptotically fractal. For details see references \cite{Meakinbook,Amitrano}. The radius of gyration $r_g$ is defined as
\begin{equation}
r_g=\left(\frac{1}{N}\sum_{i=1}^Nr_i^2\right)^{\frac{1}{2}},
\end{equation} 
where $r_i$ is the distance from the origin of the $i$th aggregated particle and $N$ the total number of particles in the cluster. The gyration radius scales with the number of particles as
\begin{equation}
r_g=BN^\nu,
\end{equation}
where $\nu=1/d_f$ and $B$ can be related to the lacunarity trough the relation
\begin{equation}
A_g=\frac{1}{B^{1/\nu}}.
\end{equation}
Notice that $A_m\ne A_g$ since their definitions are different. But, they exhibit the same qualitative behavior because $A_m\propto A_g$. 

\begin{figure*}[hbt]
\begin{center}
\includegraphics[clip=true,width=7cm,height=!]{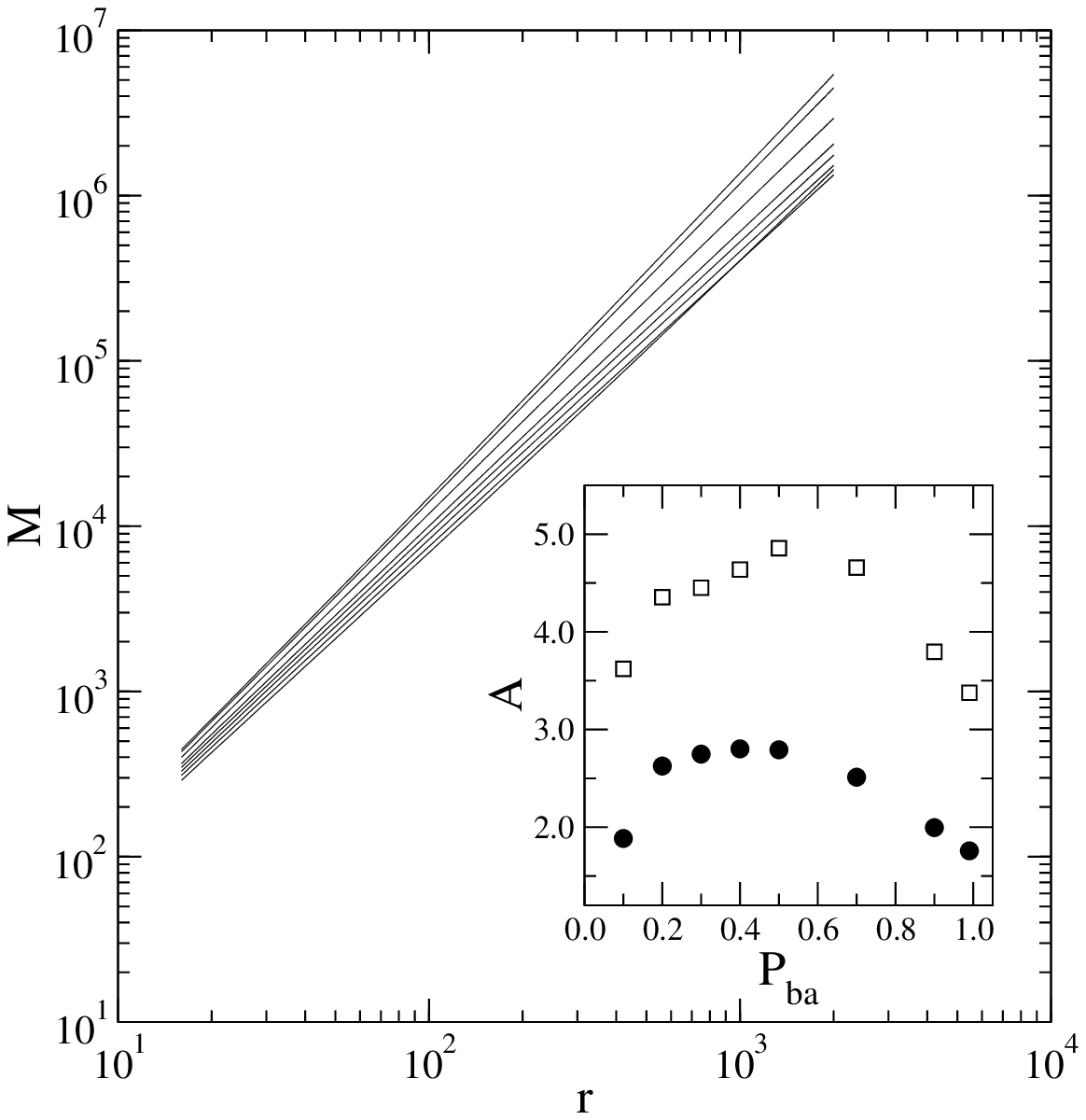}
\includegraphics[clip=true,width=7cm,height=!]{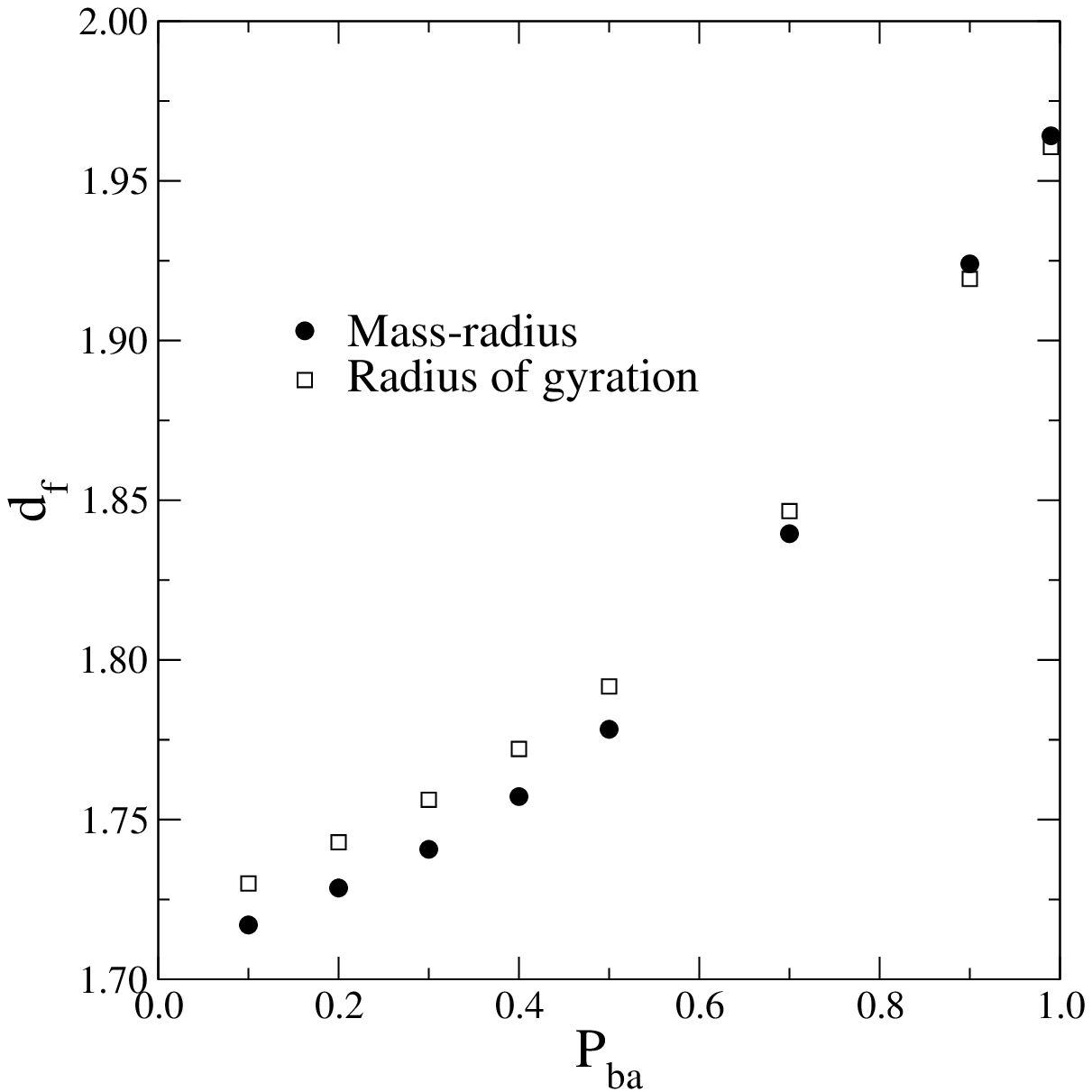}
\caption{\label{fig:mass} (a) Mass against radius for distinct $P_{ba}$ values: $0.10,\;0.20,\;0.30,\;0.40,\;0.50\;0.70,\;0.90$, and $0.99$ from the bottom to the top. (b) Fractal dimension as a function of $P_{ba}$ determined from the mass-radius ($\bullet$) and gyration radius ($\square$) methods. In inset, we show the lacunarity as a function of $P_{ba}$ obtained from mass-radius ($\bullet$) and radius of gyration ($\square$) methods. For these simulations we used clusters of size $L=10^4$ and averages done over $10^3$ independent samples. {The number of particles in these clusters varies from $7\times 10^6$ for $P_{ba}=0.1$ to $3\times 10 ^7$ for $P_{ba}=0.99$.}}
\end{center}
\end{figure*}

In figure \ref{fig:mass}(a), the double logarithm plots of the curves $M(r)$ against $r$ are shown for several $P_{ba}$ values. The slope continuously increases from $d_f=1.723\pm 0.007$ for $P_{ba}=0.10$ (a value close to the well-known DLA fractal dimension $d_f=1.715\pm0.004$ \cite{Tolman}) towards $2.0$ as $P_{ba}$ increases, as can be seen in figure \ref{fig:mass}(b), in which the fractal dimension is plotted as a function of $P_{ba}$. Remarkably, instead of the qualitative behavior observed for the fractal dimension (an increasing function of $P_{ba}$), the lacunarity is a convex function of $P_{ba}$ with a maximum at $P_{ba} \approx 1/2$, as shown in inset of figure \ref{fig:mass}(a). These results show that if the number of DLA particles is larger than the number of BA ones, an increasing $P_{ba}$ produces patterns with more dense branches (larger $d_f$ values) but with larger empty regions (larger lacunarity). However, when the number of BA particles exceeds the number of DLA ones, the width of the branches increases while the size of empty regions decreases with $P_{ba}$. An explanation for such a behavior can be obtained from the active zone analysis described next.

We define the active zone as the fraction $f$ of the latest particles added to the cluster, and its width $\sigma_f$ as the standard deviation of the distances from the origin of these particles. So,
\begin{equation}
\sigma_f^2 = \frac{1}{fN}\sum_{i=N_f}^{N}\left(r_i-\langle r\rangle\right)^2,
\end{equation}
where $N_f=(1-f)N$ and
\begin{equation}
 \langle r\rangle = \frac{1}{fN}\sum_{i=N_f}^{N}r_i.
\end{equation}

\begin{figure}[hbt]
\begin{center}
\includegraphics[clip=true,width=8.0cm,height=!]{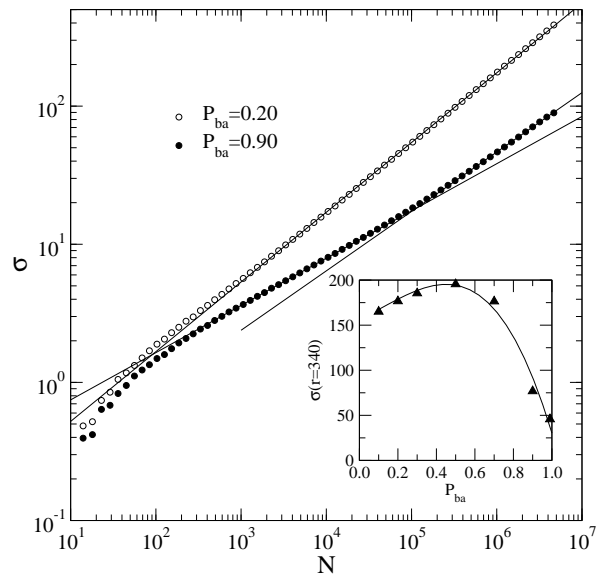}
\end{center}
\vspace{-0.5cm}
\caption{\label{fig:w_ps} Active zone width $\sigma$ as a function of the number of particles for $P_{ba}=0.20 ~(\bullet)$ and $0.90~(\circ)$. The straight lines correspond to power law fits. In inset, we show $\sigma$ as a function of $P_{ba}$ for a mean active zone radius $r\approx 340$. The line is a cubic fit used as a guide to the eyes. For these simulations we used clusters of size $L=10^4$ and averages were done over $10^3$ independent samples.}
\end{figure}

In all simulations we used $f=0.1$. In figure \ref{fig:w_ps}, the width of active zone as function of the number of particles is shown for two distinct $P_{ba}$ values. For $P_{ba}<0.5$, the simulations indicate that there is a single power law regime $\sigma\sim N^\zeta$ while a crossover between two power laws $\sigma\sim N^{\zeta'}$ for small length scales and $\sigma\sim N^{\zeta}$ for large ones is observed for $P_{ba}\ge 0.5$. In the asymptotic regime ($N\rightarrow \infty$), the exponent $\zeta \rightarrow \nu$ for $p<1$ as observed in the DLA model \cite{Tolman}. For $P_{ba} \ge 0.5$, we found $\zeta'\approx 0.46,\;0.42$ and $0.36$ for $P_{ba}=0.5,\;0.7$ and $0.9$, respectively. Indeed, our simulation suggests that $\zeta'\rightarrow 0.33$ as $P_{ba}\rightarrow 1$, in agreement with the statement that interface scaling of the BA belongs to the Kardar-Parisi-Zhang universality class  \cite{KPZ}. To be more precise, the interface scaling of ballistic deposition models were widely studied and rigorous simulations confirmed that $\sigma\sim N^{1/3}$ \cite{Meakinbook,Barabasibook}. 

A very interesting feature emerges from the behavior of $\sigma$ as a function of $P_{ba}$ for a fixed mean radius of the active zone, as shown in inset of figure \ref{fig:w_ps}. The active zone width is a convex function of $P_{ba}$ with a maximum at $P_{ba}=1/2$, which explains the lacunarity dependence on $P_{ba}$. The onset of this qualitative behavior can be understood through an analysis of the screening properties of the clusters: for  $P_{ba}\gtrsim 0$, the patterns have large grooves and the ballistic trajectories can reach the inner regions of the cluster forbidden for random walks. Thus, the screening is reduced in relation to the original DLA model and, consequently, the active zone is enlarged. As $P_{ba}$ increases, more particles break the screening enlarging the active zone. However, if too many particles arrive at the inner regions of the clusters, the size of the their grooves and, consequently, their active zones are reduced. Therefore, the convexity of $\sigma$ versus $P_{ba}$ was elucidated. Intriguingly, this change of behavior seems to occur when the number of DLA and BA particles are equal, but we cannot provide a rigorous justification besides the numerical evidence.

\begin{figure}[hbt]
\begin{center}
\includegraphics*[clip=true,width=8.0cm,height=!]{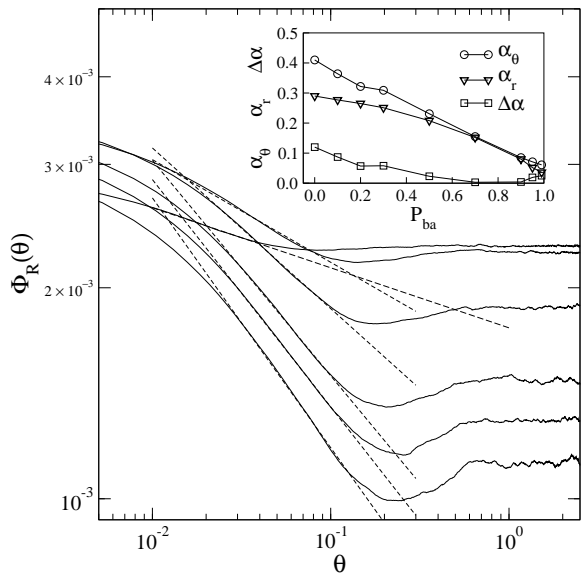}
\end{center}
\vspace{-0.5cm}
\caption{\label{fig:ang} Two point angular correlation functions evaluated at $R=300$ for several values of $P_{ba}$: $0.1$,  $0.2$, $0.3$,  $0.5$, $0.7$,  and $0.9$ from the bottom to the top, respectively.  The dashed lines correspond to the power law fits for small $\theta$. In inset, the angular and radial correlation function exponents are compared. The size of the clusters is $5000\times 5000$ and $\Phi_R$ was evaluated using $R=300$ and $\delta R = 15$. The averages were done over $N=10^3$ independent samples.}
\end{figure}

{One of the most complex features of DLA clusters is their multiscaling structures \cite{Mandelbrot, Mandelbrot1995, Ball2003}, which was recently analyzed by Mandelbrot \textit{et al.} at the light of the angular gap distributions of the clusters \cite{Mandelbrot}. They found that  the angular and radial distributions of mass scale with the same exponent $D\approx 1.71$ only for very small angular gaps. This happens for asymptotically large clusters ($M\approx 10^8$) when the number of main branches increases leading to a more uniform space fulfillment. From figure \ref{fig:pads}, one can see that angular gap distributions vary with the parameter $P_{ba}$. Therefore, a angular scaling analysis becomes necessary in order to obtain a more elaborate characterization of model. For this task, we followed the work of}
Meakin and Vicsek \cite{Meakin1985}, {in which they} analyzed the angular correlations of the DLA model using the following procedure. A narrow circular shell of radius $R$ and width $\delta R\ll R$ centered at origin is divided in $K$ identical sectors of area $R\delta\theta\delta R$, where $\delta \theta=2\pi/K$. For a box centered at the polar coordinates $(R,\phi)$, the quantity $\rho_R(\phi)=\kappa$ is defined, where $\kappa$ is the number of particles within this box. The two-point angular correlation function is given by 
\begin{equation}
\Phi_R(\theta)=\frac{1}{N}\sum_{n=0}^{K/2}\rho_R(\theta+n\delta\theta)\rho_R(n\delta\theta).
\end{equation}
The exponent obtained for the {relatively small DLA clusters ($\sim 10^5$)} using this procedure was $\alpha_\theta\approx 0.41$, which is definitely different from $\alpha_r\approx 0.29$.

\begin{figure}[hbt]
\begin{center}
\includegraphics*[clip=true,width=8.0cm,height=!]{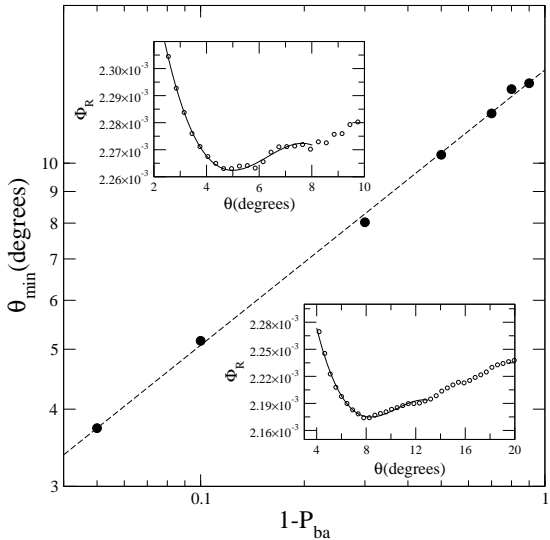}
\end{center}
\vspace{-0.5cm}
\caption{\label{fig:thetamin} Behavior of the minima presented in angular correlation functions for distinct probabilities $P_{ba}$. The dashed line is a power law fit corresponding to an exponent $\vartheta \approx 0.45$. Data in the range $P_{ba}\in [0.1,0.95]$ are shown. The insets show the linear plots of $\Phi_R$ for the regions near to the minima for $P_{ba}=0.7$ (bottom) and $P_{ba}=0.9$ (top). The solid lines are the cubic fits used to determine the positions of the minima.}
\end{figure}

Figure \ref{fig:ang} illustrates the angular correlation functions evaluated at $R=300$ for distinct $P_{ba}$ values. In all curves, $\Phi_R(\theta)$ decays as a power law for small angles followed by a global minimum and reaches a constant value for large angles. This behavior qualitatively reproduces the finds by Meakin and Vicsek \cite{Meakin1985}, but the power law exponents decay with $P_{ba}$ as illustrated in inset of figure \ref{fig:ang}, in which $\alpha_\theta$  and $\alpha_r$, as well as, the difference between them $\Delta\alpha$ are plotted as functions of $P_{ba}$. The exponent $\alpha_r$ was determined using the relation $\alpha_r=2-\overline{d_f}$, where $\overline{d_f}$ is the mean fractal dimension obtained through the mass-radius and radius of gyration methods (figure \ref{fig:mass}(b)). The difference $\Delta\alpha \approx 0.12$ observed for the DLA model decreases with $P_{ba}$. The minima of the curves {are interesting quantities since they represent estimates} of the half angular separation between two consecutive branches, the gaps studied by Mandelbrot {\it et al.} \cite{Mandelbrot}. {The minima position shift} to small angles and become less evident with increasing $P_{ba}$. Indeed, the extrapolation of minimum positions suggests that their values vanishes at $P_{ba}=1$ following a power law $\theta_{min}\sim (1-P_{ba})^\vartheta$, where $\vartheta\approx 0.45$, as shown in figure \ref{fig:thetamin}. The minima were determined using a cubic fit around their positions, as indicated in insets of figure \ref{fig:thetamin}. {This power law relation means that the angular gaps are present, even that small, for any $P_{ba}\ne 1$. However, one can resolve from inset of figure \ref{fig:ang}, the difference $\Delta \alpha$ vanishing for $P_{ba}\gtrapprox 0.5$. Thid behavior of $\Delta \alpha$ agrees with the concluding claim of Mandelbrot \textit{et al.} \cite{Mandelbrot} concerning the asymptotic scaling of ordinary off-lattice DLA clusters. Indeed, in our model the gaps are reduced with increasing $P_{ba}$ and, consequently, we observed the same result of reference \cite{Mandelbrot}, but without considering simulations with more than $10^8$ particles.}

\section{Summary}
\label{sec:conclusions}

In this work, we analyzed an aggregation model in which two kinds of particles are considered: a given fraction following Brownian motions and the remaining following ballistic trajectories. The model, which includes diffusion-limited aggregation (DLA) and ballistic aggregation (BA) models as limit cases, is controlled by a probability $P_{ba}$ that correspond to the fraction of BA particles which constitutes the aggregate. Large scale simulations and careful statistical sampling were used to characterize the radial and angular scaling of off-lattice aggregates. Specifically, we measured the fractal dimension $d_f$, the lacunarity, the width of active (growing) zone $\sigma$, and the angular correlation function $\Phi_R$ of clusters grown for distinct $P_{ba}$ values.

The fractal dimension $d_f$ continuously increases from $d_f=1.715\pm 0.004$ for $P_{ba}=0$ (DLA fractal dimension) to $d_f\approx 2$ (BA fractal dimension). However, the lacunarity and the width of the active zone for a fixed radius are convex functions of the probability $P_{ba}$ with maxima at $P_{ba} \approx 1/2$. The convexity of the lacunarity is directly related to the convexity of the active zone width, which in turn, was explained by the screening impairing the wandering particles to reach the inner regions of the clusters. 

The angular correlation function $\Phi_R(\theta)$ evaluated at $R=300$ exhibits a power law decay for small angles with a exponent that is a decreasing function of $P_{ba}$. The difference between angular and radial exponents observed for the original off-lattice DLA model \cite{Meakin1985} was found, but it rapidly vanishes for $P_{ba}>1/2$. Also, the curves $\Phi_R$ versus $\theta$ exhibit global minima representing the half angular separations between two consecutive branches. The simulations show that the position of the minima vanishes as $P_{ba}\rightarrow 1$ following a power law  $\theta_{min}\sim (1-P_{ba})^\vartheta$ with $\vartheta\approx 0.45$. These results indicate that the patterns are asymptotically homogeneous when $P_{ba}\rightarrow 1$. In summary, in this work we show how the scaling properties of the aggregates can be controled by varying the concentration of the kinds of wandering particles involved in the growth process. {Finally, it is important to emphasize that these results are in agreement with the angular gap distribution analysis performed by Mandelbrot \textit{et al.} \cite{Mandelbrot}, in which they found that the original DLA clusters exhibit the same asymptotic angular and radial scaling.}

\begin{acknowledgments}
We thank to M. L. Martins for the critical reading of this manuscript. This work was partially supported by CNPq and FAPEMIG Brazilian agencies through projects and undergraduate scholarships.
\end{acknowledgments}


\begin{thebibliography}{99}

\bibitem{Meakinbook} P. Meakin, \textit{Fractals, scaling and growth far from equilibrium} (Cambridge University Press, Cambridge, 1998).

\bibitem{Sander} L. M. Sander, Contemp. Phys. \textbf{41}, 203 (2000).

\bibitem{Witten} T. A. Witten and L. M. Sander, Phys. Rev. Lett. \textbf{47}, 1400 (1981).

\bibitem{Somfai} E. Somfai, L. M. Sander, and R. C. Ball, Phys. Rev. Lett. \textbf{83}, 5523 (2000).

\bibitem{Mandelbrot} B. B. Mandelbrot, B. Kol, and A. Aharony, Phys. Rev. Lett. \textbf{88}, 055501 (2002).

\bibitem{Jensen} M. H. Jensen, J. Mathiesen and I. Procaccia, Phys. Rev. E \textbf{67}, 042402 (2003).

\bibitem{Goold} N. R. Goold, E. Somfai, R. C. Ball, Phys Rev. E \textbf{72}, 031403 (2005).

\bibitem{Vold} M. J. Vold, J. Colloid Sci. \textbf{18}, 684 (1963).

\bibitem{Liang} S. Liang and L. P. Kadanoff, Phys. Rev. A \textbf{31}, 2628 (1985).

\bibitem{Ferreira2005} S. C. Ferreira Jr., S. G. Alves, A. F. Brito, and J. G. Moreira, Phys. Rev. E \textbf{71}, 051402 (2005).

\bibitem{Castro} M. Castro, R. Cuerno, A. S\'anches, and F. Dom\'{\i}gues-Adame, Phys. Rev. E \textbf{62}, 161 (2002).

\bibitem{Kim} Y. Kim, K. R. Choi, and H. Pak, Phys Rev. A \textbf{45}, 5805 (1992). 

\bibitem{Meakin1983} P. Meakin, Phys. Rev. B \textbf{28}, 5221 (1983).

\bibitem{Nagatani} T. Nagatani, Phys. Rev. A \textbf{39}, 438 (1989). 

\bibitem{Meakin1985} P. Meakin and T. Vicsek, Phys. Rev. A \textbf{32}, R685 (1985).

\bibitem{Mandelbrot1995} {B. B. Mandelbrot, A. Vespignani, and H. Kaufman, Euro. Phys. Lett. \textbf{32}, 199 (1995).}

\bibitem{Ball2003} {E. Somfai, R. C. Ball., N. E. Bowler, and L. M. Sander, Physica A \textbf{325}, 19 (2003).}

\bibitem{Tolman} S. Tolman and P. Meakin, Phys. Rev. A \textbf{40}, 428 (1989).

\bibitem{Tang} C. Tang and S. Liang, Phys. Rev. Lett. {\bf 71}, 2769 (1993).

\bibitem{Yu} J. Yu and J. G. Amar, Phys. Rev. E {\bf 66}, 021603 (2002).

\bibitem{Ball} R. C. Ball and R. M Brady, J. Phys. A: Math. Gen. \textbf{18}, L809 (1985).

\bibitem{Amitrano} C. Amitrano, A. Coniglio, P. Meakin, and M. Zannetti, Phys. Rev. B \textbf{44}, 4974 (1991).

\bibitem{KPZ} M. Kardar, G. Parisi, and Y. C. Zhang, Phys. Rev. Lett. \textbf{56}, 889 (1986).

\bibitem{Barabasibook} A. -L. Barabasi and E. H. Stanley, \textit{Fractal Concepts on Surface Growth} (Cambridge University Press, Cambridge, 1995).

\end{thebibliography}
\end{document}